\documentclass[preprint,superscriptaddress,notitlepage,endfloat]{revtex4-1}
\usepackage{graphicx,bm,mathtools,color}

\begin{document}

\renewcommand{\abstractname}{} 
\title{Evidence for Weyl fermions in a canonical heavy-fermion semimetal YbPtBi}

\author{C. Y. Guo}
\affiliation{Center for Correlated Matter and Department of Physics, Zhejiang University, Hangzhou 310058, China}
\author{F. Wu}
\affiliation{Center for Correlated Matter and Department of Physics, Zhejiang University, Hangzhou 310058, China}
\author{Z. Z. Wu}
\affiliation{Center for Correlated Matter and Department of Physics, Zhejiang University, Hangzhou 310058, China}
\author{M. Smidman}
\affiliation{Center for Correlated Matter and Department of Physics, Zhejiang University, Hangzhou 310058, China}
\author{C. Cao}
\affiliation{Department of Physics, Hangzhou Normal University, Hangzhou 310036, China}

\author{A. Bostwick}
\affiliation{Advanced Light Source, E.O. Lawrence Berkeley National Lab, Berkeley, California 94720, United States}
\author{C. Jozwiak}
\affiliation{Advanced Light Source, E.O. Lawrence Berkeley National Lab, Berkeley, California 94720, United States}
\author{E. Rotenberg}
\affiliation{Advanced Light Source, E.O. Lawrence Berkeley National Lab, Berkeley, California 94720, United States}
\author{Y. Liu}
\affiliation{Center for Correlated Matter and Department of Physics, Zhejiang University, Hangzhou 310058, China}
\author{F. Steglich}
\affiliation{Center for Correlated Matter and Department of Physics, Zhejiang University, Hangzhou 310058, China}
\affiliation{Max Planck Institute for Chemical Physics of Solids, 01187 Dresden, Germany}
\author{H. Q. Yuan*}
\affiliation{Center for Correlated Matter and Department of Physics, Zhejiang University, Hangzhou 310058, China}
\affiliation{Collaborative Innovation Center of Advanced Microstructures, Nanjing University, Nanjing 210093, China}

\date{\today}
\begin{abstract}
\textbf{The manifestation of Weyl fermions in strongly correlated electron systems is of particular interest. We report evidence for Weyl fermions in the heavy fermion semimetal YbPtBi from electronic structure calculations, angle-resolved photoemission spectroscopy, magnetotransport and calorimetric measurements. At elevated temperatures where $4f$-electrons are localized, there are triply degenerate points, yielding Weyl nodes in applied magnetic fields. These are revealed by a contribution from the chiral anomaly in the magnetotransport, which at low temperatures becomes negligible due to the influence of electronic correlations. Instead, Weyl fermions are inferred from the topological Hall effect, which provides evidence for a Berry curvature, and a cubic temperature dependence of the specific heat, as expected from the linear dispersion near the Weyl nodes. The results suggest that YbPtBi is a Weyl heavy fermion semimetal, where the Kondo interaction renormalizes the bands hosting Weyl points. These findings open up an opportunity to explore the interplay between topology and strong electronic correlations.}
\end{abstract}

\maketitle
The presence of topologically non-trivial  electronic band structures  in condensed matter systems leads to a number of unusual consequences \cite{RMPTopo}. A  rich variety of phenomena have  been discovered in gapless topological materials, such as those exhibiting Dirac-fermion excitations near the points of linear crossings of bands close to the Fermi energy $E_{\rm F}$  \cite{Dirac1,Dirac2}. The breaking of either spatial inversion symmetry or time reversal symmetry splits the degeneracy of the Dirac points, leading to a pair of topologically protected  Weyl points \cite{Weyl1,Weyl2}. Weyl fermions have been found to cause distinct experimental signatures such as the chiral anomaly in transport measurements \cite{Nielsen1983,ChiralAnomTheor1,WeylTaAsChirAnom1}, a topological Hall effect \cite{GdPtBiAHE,nakatsuji2015large,li2018momentum} and Fermi arcs \cite{ArcRev}.

Weyl fermions have mainly been studied in weakly correlated electron systems, while strong electronic correlations are  frequently found to lead to novel electronic properties beyond those of simple metals or insulators, and heavy fermion systems are the prototype examples showing phenomena characteristic for strongly correlated electron systems. Here, due to strong  Kondo coupling between the $f$-electron and conduction-band states, below the Kondo temperature ($T_{\rm K}$), the electronic bands in the vicinity of $E_{\rm F}$ may become strongly renormalized, showing a strong $f$-character and a huge enhancement of the quasiparticle mass. When the chemical potential lies within the hybridization gap, insulating behavior is found at low temperatures and in the topological Kondo insulators, such as has been proposed for SmB$_6$, the resulting electronic structure is topologically non-trivial, again leading to conducting states on the surface \cite{SmB6a,SmB6f,SmB6c}. It is therefore of particular interest to look for topological heavy fermion semimetals with gapless excitations, i.e. Weyl fermions in the presence of strongly renormalized bands. Such a Weyl-Kondo semimetal phase has been predicted from calculations based on the periodic Anderson model with broken inversion symmetry \cite{SiHFEeyl,Chang2018}. While it was proposed that Ce$_3$Bi$_4$Pd$_3$ displays the low temperature thermodynamic signatures of a Weyl-Kondo semimetal \cite{SiHFEeyl,dzsaber2016tuning},  other signatures of Weyl fermions such as the chiral anomaly have not been reported. A Weyl heavy fermion state was also proposed for CeRu$_4$Sn$_6$ from ab initio calculations \cite{CeRu4Sn6Theor}, but no experimental evidence for Weyl fermions has been demonstrated.   Consequently, whether Weyl fermions exist in the presence of a strong Kondo effect needs to be determined experimentally.  Furthermore,  the influence of electronic correlations on Weyl fermions is to be explored, specifically how such a system evolves from high temperatures, where the $f$-electrons are well localized, to low temperatures where there is a strong Kondo interaction and  a reconstruction of the electronic bands.

The cubic half Heusler compounds (space group $F\bar{4}3m$) can be tuned by elemental substitution from trivial to topological insulators \cite{Canfield1991RBiPt,chadov2010tunable}. It was recently found that the half Heusler  GdPtBi, which has a strongly localized $4f$-electron shell, shows evidence for Weyl fermions in an applied magnetic field due to the presence of the chiral anomaly \cite{GdPtBiChiral} and topological Hall effect  \cite{GdPtBiAHE}. Here we examine the isostructural compound YbPtBi. Although at high temperatures the Yb $4f$-electrons are localized similar to GdPtBi, upon cooling YbPtBi becomes a prototypical heavy-fermion semimetal \cite{YbBiPt1991,Hundley1997,TwoChanYbPtBi}, where the enormous Sommerfeld coefficient of $\gamma~\approx~8$~J~mol$^{-1}$~K$^{-2}$  demonstrates the enhanced effective mass of the charge carriers \cite{YbBiPt1991}. This compound is therefore highly suited to look for Weyl fermions which are strongly affected by electronic correlations. 

In this work, we report evidence for Weyl fermions in YbPtBi, where the bands hosting the Weyl points are strongly modified as the Kondo coupling strengthens at low temperatures. Electronic structure calculations and angle resolved photoemission spectroscopy (ARPES) measurements indicate the presence of triply degenerate fermion points in the high temperature regime, which will each split into a Weyl node and a trivial crossing in applied fields. At these temperatures, evidence for the chiral anomaly is revealed by field-angle dependent magnetotransport measurements. As the temperature is lowered, the chiral anomaly is not detected in the magnetotransport, but experimental signatures of Weyl fermions are found in measurements of the specific heat. This is consistent with a greatly reduced Fermi velocity due to the influence of the Kondo effect on the electronic bands near the Weyl points. Furthermore, the observation of a topological Hall effect contribution, which can arise from the Berry curvature generated by the Weyl nodes, provides additional evidence for the existence of Weyl fermions at both low and elevated temperatures.

\section*{Results}

\textbf{ARPES and electronic structure calculations.} At higher temperatures, the band structure of YbPtBi can be calculated  treating $f$-electrons as core states, as displayed in Fig.~1. The bulk Fermi surface consists of hole pockets centered at the $\Gamma$-point and electron pockets slightly away from $\Gamma$. Along $\Gamma$-L, the four-fold degenerate $\Gamma_8$ state splits into two non-degenerate hole bands, and a pair of degenerate $\Lambda_6$ electron  bands, mainly consisting of Yb-$t_{\rm 2g}$ and Bi-$p$ orbitals. The  $\Lambda_6$  bands cross the two hole bands near $E_{\rm F}$, forming two triply degenerate fermion points \cite{lv2017observation}. Under a magnetic field, each triply degenerate point will further split into a Weyl point and a trivial crossing, with energies close to the bottom of the electron bands. The calculated bulk band structure with triply degenerate points is in good agreement with the ARPES results in Fig. 1b, which shows the energy-momentum dispersion relations along the surface $\bar{\Gamma}\bar{M}$ direction. Note that the sample can only be  cleaved well with the (111) orientation. Along this orientation, the symmetry-equivalent bulk  $\Gamma L$ direction projects on the surface $\bar{\Gamma}\bar{M}$ direction at a slanted angle, allowing for the  dispersion in the vicinity of the triply degenerate points to be revealed via a careful comparison with the projected bulk band structure calculations (Fig.~1c). Two hole bands crossing $E_{\rm F}$ can be clearly identified in the ARPES experiments, as well as an additional electron band with a band bottom right below $E_{\rm F}$. These experimentally observed bands are confirmed to be three-dimensional bulk bands based on their photon energy dependence, and they correspond well to the theoretical calculations. The direct observation of both electron and hole pockets and their close proximity with different group velocities confirms the existence of the triply degenerate fermion points near $E_{\rm F}$, which is not affected by the slight discrepancy between the experimental results and calculations. This discrepancy is mainly related to the details of the separation and slope of the two hole bands, which could be caused by the limitations of frozen $f$-shell calculations and correlation effects not taken into account by the local density approximation. The good correspondence between ARPES measurements and density functional theory (DFT)  calculations therefore provides evidence for Weyl fermions at elevated temperatures.  \\

\noindent\textbf{Probing the chiral anomaly using magnetotransport.} Magnetotransport measurements were performed to look for the chiral anomaly associated with Weyl fermions (Fig.~2). Figures~2a-2d show the field dependence of the resistivity of YbPtBi at selected temperatures with a current $\mathbf{I}$  along [100] and a magnetic field $\mathbf{B}$ applied parallel and perpendicular to $\mathbf{I}$.  For temperatures between 25~K and 170~K, the longitudinal magnetoresistance ($\mathbf{B}\parallel\mathbf{I}$) is positive at low fields but becomes negative in the higher field region, while the transverse magnetoresistance ($\mathbf{B}\perp\mathbf{I}$) is positive, which together are evidence for the chiral anomaly. The negative longitudinal magnetoresistance cannot be explained by either current jetting  (Supplementary Fig.~2 and Supplementary Note~1) \cite{Currentjet}, nor the sample anisotropy since similar behavior is found for other current directions  (Supplementary Fig.~3). The negative longitudinal magnetoresistance above 20 K could be well fitted  using a conductivity  $\sigma(B)=(1+c_{\rm a}B^2)\sigma_{\rm WAL}$ (Figs. 2a-2c, Supplementary Fig.~4 and Supplementary Note~2), where $c_{\rm a}$ is the chiral constant and $\sigma_{\rm WAL}=\sigma_{\rm N}+a\sqrt{B}$  is due to the weak antilocalization \cite{ChiralAnomTheor1,WeylTaAsChirAnom1}. As shown in Fig.~2e, the temperature dependence of $c_{\rm a}$ is well fitted with the expected behavior of $c_{\rm a}\propto v_{\rm F}^3\tau_{\rm v}/(T^2+\mu^2/\pi^2)$, where $\tau_{\rm v}$ is the chirality-changing scattering time and $\mu$ is the chemical potential \cite{ZrTe5Chiral}, yielding $v_{\rm F}^3\tau_{\rm v}~=~134$~m$^3$s$^{-2}$ and $\mu=1.5$~meV.  $\sigma(B)$ for various  angles $\theta$ between $\mathbf{B}$ and $\mathbf{I}$ are displayed in Figs.~2f-2h as a function of $B^2$, where the high field linear behavior  indicates a $B^2$ contribution, while the very small $a$ values lead to a negligible component $\propto B^{\frac{5}{2}}$ (Fig.~2e). As displayed in Fig.~2i (and Supplementary Fig.~5), the extracted $c_{\rm a}(\theta)$ shows the  expected  angular dependence of $c_{\rm a}(\theta)\sim\rm{cos}^2\theta$. Therefore both the angle and temperature dependence of the magnetoresistance are highly consistent with the presence of a chiral anomaly in YbPtBi.

Meanwhile either by changing the Bi flux concentration or by Au doping, the carrier concentration can be tuned, as shown in Fig.~3. The Hall resistivity for various samples shows that  more strongly hole doped samples exhibit  one band behavior with larger hole densities ($n_{\rm H}$), but upon electron doping, $E_{\rm F}$ is shifted and eventually crosses the electron bands, leading to two band behavior (Fig.~3a, Supplementary Figs.~6 and 7). As shown in Figs.~3b and 3c,  in the vicinity of the crossover between one and two band behavior, the negative longitudinal magnetoresistance is most prominent. For more strongly electron or hole doped samples, no negative magnetoresistance is seen at elevated temperatures, indicating that this negative longitudinal magnetoresistance arises when  $E_{\rm F}$ is close to the Weyl points (Figs. 1a and 3d). Measurements of the transverse resistivity (with the voltage measured perpendicular to  $\mathbf{I}$) for fields rotated in the plane of the voltage drop and  $\mathbf{I}$ ($\rho_{\rm xy}^{\rm PAMR}$) provide an alternative method for probing the chiral anomaly which is much less sensitive to spin scattering than the magnetoresistance  (Figs.~3e-3h) \cite{PHEGdPtBi,PHEGdPtBi2}. For two samples with evidence for the chiral anomaly in the magnetoresistance (S7 and S9), the oscillation amplitude of $\rho_{\rm xy}^{\rm PAMR}$ is greatly enhanced above 20 K, while this remains small for the more electron-doped sample, which is another signature of the chiral anomaly in samples where $E_{\rm F}$ is near the band crossing. Interestingly, at 2~K the oscillations have very small amplitudes and are not sample dependent (Fig.~3f). This suggests that evidence for the chiral anomaly disappears from these measurements at low temperatures, leaving only  a small  contribution likely from the orbital magnetoresistance. Similar conclusions are drawn  from the magnetoresistance in Fig.~2d, which at low temperatures is negative at \textit{all} $\theta$, and the behavior is well accounted for by single impurity Kondo scaling  \cite{schlottmann1989some} (Supplementary Fig.~9 and Supplementary Note~5). This disappearance may be related to the drop of the effective Fermi velocity to $v^*\ll v_{\rm F}$ as the quasiparticles gain mass in the heavy fermion state, since $c_{\rm a}\propto v_{\rm F}^3$ and therefore  decreasing $v_{\rm F}$ will greatly reduce the chiral anomaly contribution. As a result, the disappearance of the chiral anomaly at low temperatures suggests a significant modification of the Weyl points by the electronic correlations.\\

\noindent\textbf{Topological Hall effect.} Even in the case when $v_{\rm F}$ is small, the Berry curvature induced by the Weyl points can still contribute to the anomalous Hall effect (AHE) \cite{Burkov2011}. We analyzed the Hall resistivity between 0.3 and 30~K by considering the total Hall resistivity as the sum of three terms \cite{GdPtBiAHE,nakatsuji2015large,Li2013}

\begin{equation}
\rho_{\rm xy} = \rho_{\rm xy}^{\rm N} + \rho_{\rm xy}^{\rm A} + \rho_{\rm xy}^{\rm T}
\end{equation}
 where $\rho_{\rm xy}^{\rm N}$,  $\rho_{\rm xy}^{\rm A}$, and  $\rho_{\rm xy}^{\rm T}$ are the normal Hall effect, anomalous term from the magnetization, and the topological Hall effect term arising from the Berry curvature, respectively \cite{GdPtBiAHE,HWBRev}. Figure~4a shows the anomalous contribution to the Hall resistivity ($\rho_{\rm xy}^{\rm A}+\rho_{\rm xy}^{\rm T}$) after subtracting the ordinary band part $\rho_{\rm xy}^{\rm N}$; the data are taken from measurements of sample~S6 which exhibits single band behavior and evidence for the chiral anomaly. Here the $\rho_{\rm xy}^{\rm A}$  term shown by the dashed lines is proportional to the magnetization, which dominates at higher temperatures due to an increased resistivity (Supplementary Fig.~8), while the topological part $\rho_{\rm xy}^{\rm T}$ gives rise to the maxima in Fig.~4a at low temperatures. After subtracting $\rho_{\rm xy}^{\rm A}$, the topological Hall angle  $\Theta_{\rm xy}^{\rm T}=\Delta\sigma_{\rm xy}^{\rm T}/\sigma_{\rm xx}$ is obtained and is displayed in  Fig.~4b. Here a peak  in $\Theta_{\rm xy}^{\rm T}$ can be resolved up to temperatures of at least 30~K, which is very similar to the behavior observed in the magnetic Weyl semimetals GdPtBi \cite{GdPtBiAHE} and Mn$_3$Sn  \cite{li2018momentum}. The  large maximum value in $\Theta_{\rm xy}^{\rm T}$ of 0.18 at 0.3~K in YbPtBi is comparable to the respective values  of 0.17 and 0.4 for the two other compounds \cite{GdPtBiAHE,li2018momentum}. We note that in the regions where the Hall resistivity is linear (below around 0.2~T and above 4.6~T at 0.3~K), the slope of $\rho_{\rm xy}$ is very similar. This indicates that the carrier concentration does not change significantly up to the maximum measured field, and therefore the observed $\Theta_{\rm xy}^{\rm T}$ does not likely arise due to a significant change in the electronic structure. Consequently, these results provide evidence that even at low temperatures, the Berry curvature from the Weyl points is still manifested in the anomalous Hall effect.\\

\noindent\textbf{Evidence for Weyl nodes from the specific heat.} Evidence for the presence of Weyl points in the heavy fermion state is also found in specific heat measurements. While  in zero field there is an upturn of $C(T)/T$  prior to the onset of antiferromagnetic order in zero-field at 0.4~K (Supplementary Fig.~10) \cite{YbBiPt1991,Mun2013YbPtBi}, for larger applied fields  $C(T)/T$  reaches a maximum before decreasing at lower temperatures. However, as also shown by the solid lines  in Fig.~4c, the low temperature $C(T)/T$ at higher fields deviates from a spin-1/2 resonance-level model for Kondo impurity systems (Supplementary Note~6) \cite{SCHOTTE197538}, where two levels of width $\Delta$ are split by a Zeeman field. This model can be widely applied in heavy fermion systems, both in the coherent heavy Fermi liquid state and the dilute limit \cite{Pikul2012}. In higher fields, $C/T$  of the Kondo impurity model becomes nearly temperature independent at low temperatures, but the data are instead well described by a $T^3$ dependence of the specific heat, $C\sim(k_{\rm B}T/\hbar v^*)^3$ [Fig.~4d], which was proposed for a Weyl-Kondo semimetal \cite{SiHFEeyl}, as a result of the linear dispersion $\epsilon_{\mathbf k}=\hbar v^*k$ in the vicinity of the Weyl nodes.  We note that this term is too large to arise from acoustic phonons since it would correspond to an unreasonably small Debye temperature of $\theta_{\rm D}=32$~K, compared to the much larger value of $\theta_{\rm D}=190$~K for isostructural LuPtBi \cite{Mun2013YbPtBi}. With increasing field there is  a decrease of the Sommerfeld coefficient $\gamma$  and an increase of  $v^*$, consistent with the applied field reducing the effective mass of the quasiparticles (Supplementary Table~1). However,  even at $B=13$~T a value of $\gamma=89$~mJ~mol$^{-1}$~K$^{-2}$ is obtained, indicating that a significant mass enhancement persists in this field region, which  is consistent with the single impurity Kondo scaling  present up to the maximum measured field of 9~T (Supplementary Fig.~9).   Correspondingly, fitting the data yields low effective Fermi velocities of $v^*=213$~ms$^{-1}$ at 7~T and $v^*=394$~ms$^{-1}$ at 13~T, which are significantly reduced compared to the Fermi velocity of $v_{\rm F}=2.3\times10^5$ms$^{-1}$ estimated from $v_{\rm F}=(\hbar/m_{\rm e})(3\pi^2 n_{\rm H})^{\frac{1}{3}}$ at 50~K (Fig.~3c). 

\section*{Discussion}

Based on the above experimental findings, we propose the diagram shown in Fig.~5 to describe the Weyl fermions in YbPtBi. At high temperatures there are Weyl nodes formed from the conduction  bands, while the $f$ electrons are well localized. This is consistently shown from electronic structure calculations, ARPES and magnetotransport measurements. At lower temperatures, the  strong band renormalization due to Kondo coupling enhances the effective quasiparticle mass, which modifies the dispersion of the bands in the vicinity of the topologically protected Weyl points, as shown schematically in the diagram. The renormalization also leads to a greatly reduced effective Fermi velocity $v^*$ compared to the bare band value, which eventually causes the disappearance of the chiral anomaly in transport measurements, but  allows for the observation of a sizeable specific heat contribution $C\sim(k_{\rm B}T/\hbar v^*)^3$  \cite{SiHFEeyl} . Importantly,  there is evidence for the Berry curvature associated with the Weyl nodes from the anomalous Hall effect, which can be detected in both the intermediate and low temperature regimes.

Our results highlight the existence of Weyl fermions in YbPtBi, where we find evidence for their modification as the Kondo coupling is strengthened upon lowering the temperature.  How precisely the Weyl points are modified as the electronic correlations become stronger  needs to be determined by future studies. While the topological Hall effect and specific heat provide evidence for the survival of Weyl fermions at low temperatures, looking for spectroscopic evidence from ARPES  or scanning tunneling spectroscopy is very important. One possible approach to reveal  Weyl fermions in the heavy fermion state from $f$-bands is  resonant photoemission. However, our measurements across the Yb N edge do not show obvious resonance contrast (Supplementary Fig.~11). Although ARPES measurements with $h\nu>100$~eV (including with soft x-rays) indeed reveal the bulk $f$ bands near $E_{\rm F}$ (Supplementary Fig.~12), resolving the (fine) hybridized bands deep inside the heavy fermion state is still challenging, and therefore further ARPES measurements with greater energy and momentum resolution are highly desirable.

The presence of Weyl fermions in YbPtBi is different from the cases of both CeSb \cite{guo2016cesb} and GdPtBi \cite{GdPtBiChiral}, where the  bands hosting  Weyl fermions do not have a significant $f$-electron contribution. Meanwhile, evidence for Weyl fermions has also been found in some magnetic $d$-electron systems such as Mn$_3$Sn \cite{Mn3SnWeyl} and YbMnBi$_2$ \cite{YbMnBi2a,YbMnBi2b}, where in the case of Mn$_3$Sn a significant topological Hall effect is also observed \cite{nakatsuji2015large,li2018momentum}. On the other hand, it is of great interest to look for the kind of dichotomy observed here for YbPtBi in other potential Weyl heavy-fermion semimetals, such as Ce$_3$Bi$_4$Pd$_3$ where a similarly small $v^*$ was inferred from the specific heat \cite{dzsaber2016tuning}, yet evidence for the chiral anomaly at elevated temperatures has not yet been reported. Furthermore, the strength of the Kondo interaction in heavy fermion systems can be readily tuned  by non-thermal control parameters such as pressure and magnetic field,  and in particular,  a quantum critical point can be reached in YbPtBi at a critical field of 0.4~T  \cite{Mun2013YbPtBi}. Therefore, our findings may open up the opportunity to explore the exciting relationship between Weyl fermions, electron-electron correlations and quantum criticality.

\section*{Methods} 

\noindent\textbf{Sample synthesis.}  Single crystals of YbPtBi were prepared using a Bi self flux \cite{canfield1992growth}. Elemental Yb, Pt and Bi  were combined in a range of molar ratios from 1:1:7-1:1:20 and heated to 1150$^\circ$C, before being slowly cooled to $500^\circ$C at a rate of $4^\circ$C/hour. For some samples, Au was also added up to a maximum ratio of Au:Pt of 1:19. The single crystal quality and orientation were checked using Laue diffraction, which was measured along the [100] direction (Supplementary Fig.~1). 

\noindent\textbf{Physical properties characterization.} The  magnetotransport was measured using the four-probe method in a Quantum Design Physical Property Measurement System (9T-PPMS) with the sample rotation option, where Pt wires were attached to the sample.  Hall effect measurements for determining the anomalous Hall effect were performed in a $^3$He cryostat with a 15~T magnet.  As shown in Supplementary Fig.~2, for some samples multiple voltage contacts were made, so as to rule out current inhomogeneities and the current jetting effect. The temperature dependence of the resistivity was checked for  several samples  (Supplementary Fig.~1), which are similar to previous reports \cite{Mun2013YbPtBi}. The resistivities at 2~K range from $27-37\mu\Omega$-cm with $\rho(300~{\rm K})/\rho(2~{\rm K})\approx10$. 

Specific heat measurements were performed using a 14T-PPMS using a $^3$He option, while magnetization measurements were carried out using the vibrating sample magnetometer (VSM) option. The magnetic susceptibility data are well fitted by the Curie-Weiss expression between 10 and 300~K (Supplementary Fig.~1), yielding a Curie-Weiss temperature of $\theta_{\rm CW}~=~-2.3$~K and an effective moment of $4.29\mu_{\rm B}$/Yb, again consistent with previous results \cite{Mun2013YbPtBi}.

\noindent\textbf{ARPES measurements and electronic structure calculations.} ARPES measurements, including resonant photoemission across the Yb N edge, were performed at the Advanced Light Source, BL7 micro-ARPES beamline. The (111)-oriented YbPtBi samples were cleaved in-situ and measured at around 20 K with 75~eV photons, unless noted otherwise. A detailed photon energy dependence study was carried out to confirm the bulk nature of the bands reported here. The typical domain size after cleavage is only a few tens of $\mu$m for the Yb termination. The surface termination (either Yb or Bi terminated) is determined by core level analysis, as well as a detailed comparison with DFT calculations. The soft X-ray ARPES measurements (Supplementary Fig.~12) were performed at the ID29, Advanced Photon Source. The DFT calculations were performed with plane-wave basis and projected augmented wave method as implemented in VASP. The $f$-electrons are treated as core-states in these calculations. To ensure convergence, plane-waves up to 480 eV and 12x12x12 $\Gamma$-centered K-mesh was employed. The generalized gradient approximation is known to overestimate the band inversions in crystal, therefore we have employed modified Becke-Johnson potentials to calculate the band structure. 
\\
\\
\textbf{Acknowledgments} We would like to thank Qimiao Si, Joe Thompson, Fuchun Zhang,  Jianhui Dai and Yi Zhou for valuable discussions. We also thank Dr. J. McChesney and Dr. F. Rodolakis Simoes for beamline support during the soft x-ray ARPES measurements.  This work was supported by the National Key R\&D Program of China (No.~2017YFA0303100, No.~2016YFA0300202), the National Natural Science Foundation of China (No.~U1632275, No.~11474251) and the Science Challenge Project of China (No.~TZ2016004). The ALS and APS are supported by the Office of Basic Energy Sciences of the U.S. DOE under Contract Nos. DE-AC02-05CH11231 and DE-AC02-06CH11357, including additional support by National Science Foundation under Grant no. DMR-0703406.\\
\\
\textbf{Data availability} All the data supporting the findings are available from the corresponding author upon reasonable request.
\\
\\
\textbf{Additional information} Correspondence and requests for materials should be addressed to H. Q. Yuan (hqyuan@zju.edu.cn)
\\
\\
\textbf{Author contributions} The project was conceived by C.Y.G. and H.Q.Y.. The crystals were grown  by  F.W.. Magnetotransport and specific heat measurements were performed by C.Y.G. and  F.W., and analyzed by C.Y.G., F.W., M.S., F.S., and H.Q.Y.. Electronic structure calculations were carried out by C.C.. ARPES measurements were performed and analyzed by Z.Z.W, A.B., C.J., E.R., and Y.L..    C.Y.G., M.S., F.S., C.C., Y. L. and H.Q.Y. wrote the manuscript.
\\
\\
\textbf{Competing interests} The authors declare no competing interests.

\clearpage
\begin{figure}[h]
\begin{center}
 \includegraphics[width=0.99\columnwidth]{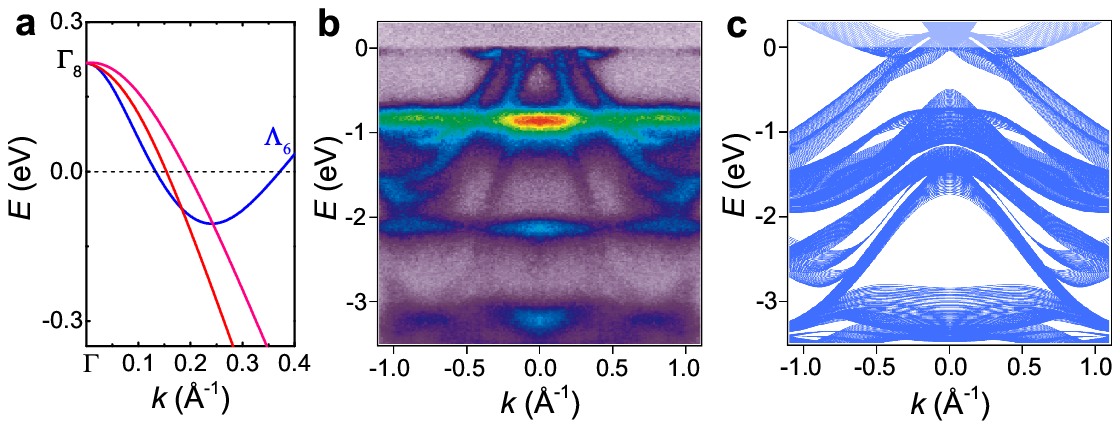}
\end{center}
	\caption{\textbf{Existence of triply degenerate fermion points in the high temperature phase of YbPtBi from DFT and ARPES.} \textbf{a},  Bulk band structure of YbPtBi from DFT calculations along the [111] direction. Blue curves are doubly degenerate $\Lambda_6$ states, while the red and cyan curves represent the non-degenerate hole states. A comparison is shown between \textbf{b}, ARPES measurements and \textbf{c}, projected bulk band structure calculations, which are in good agreement.  The comparison was made for the Yb-terminated (111) surface along the in-plane $\bar{\Gamma}-\bar{M}$ direction ($[11\bar{2}]$). The flat $f$ bands at energies of -0.9 and -2.1 eV are the surface $f$ bands from the topmost Yb layer (Supplementary Fig.~12), which are not taken into account by the bulk DFT calculations. }
   \label{Fig1}
\end{figure}

\begin{figure}[h]
\begin{center}
 \includegraphics[width=0.99\columnwidth]{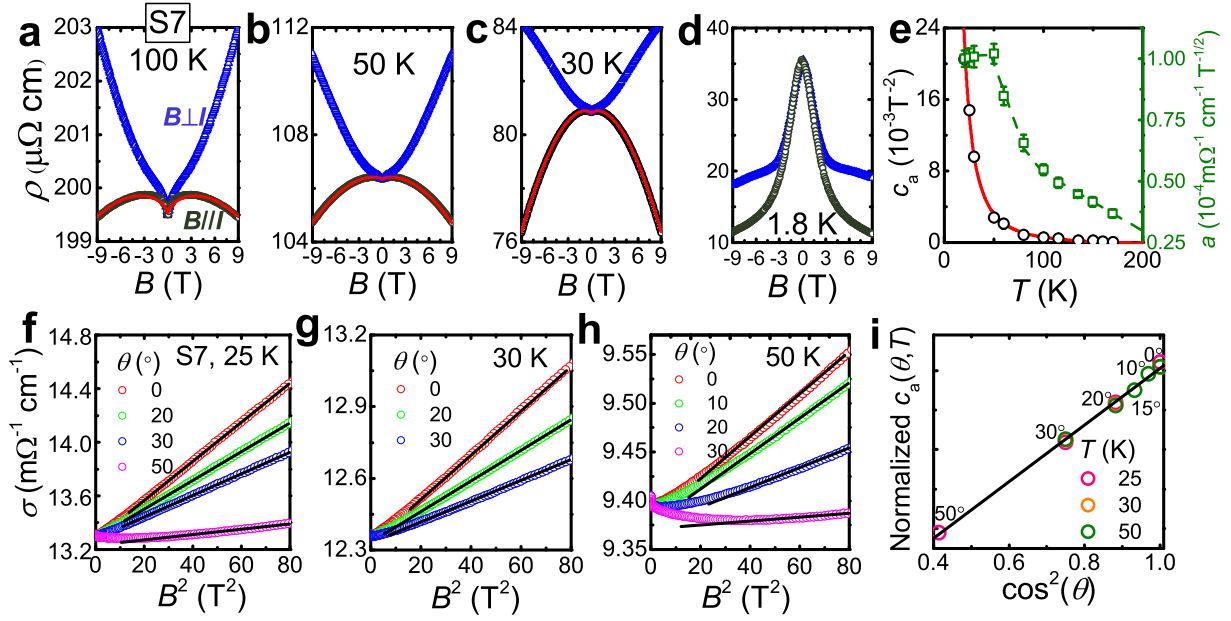}
\end{center}
	\caption{\textbf{Evidence for the chiral anomaly in YbPtBi at elevated temperatures from magnetotransport.}   Field dependence of the resistivity of YbPtBi at \textbf{a}, 100~K, \textbf{b}, 50~K, \textbf{c}, 30~K, and \textbf{d}, 1.8~K, for fields parallel and perpendicular to the current $\mathbf{I}\parallel[100]$. The solid lines show fits for  $\mathbf{B}\parallel\mathbf{I}$ at elevated temperatures, taking into account the chiral anomaly and the weak antilocalization, as described in the text. \textbf{e}, Temperature dependence of the chiral constant $c_{\rm a}$  and the  weak antilocalization  $a$-coefficient. The error bars correspond to the standard errors from the fits performed using the least squares method. The solid line shows the fitted temperature dependence of $c_{\rm a}$  \cite{ZrTe5Chiral}. The conductivity of sample S7 versus $B^2$ for various $\theta$, where $\theta$ is the angle between the field and current is shown at  \textbf{f}, 25~K, \textbf{g}, 30~K, and \textbf{h}, 50~K,   which exhibit the expected $B^2$ dependence at high fields. \textbf{i}, Normalized chiral constant $c_{\rm a}$ obtained from fitting the high field conductivity, where $c_{\rm a}$ shows a cos$^2\theta$ dependence. }
   \label{Fig2}
\end{figure}

\clearpage
\begin{figure}[h]
\begin{center}
 \includegraphics[width=0.99\columnwidth]{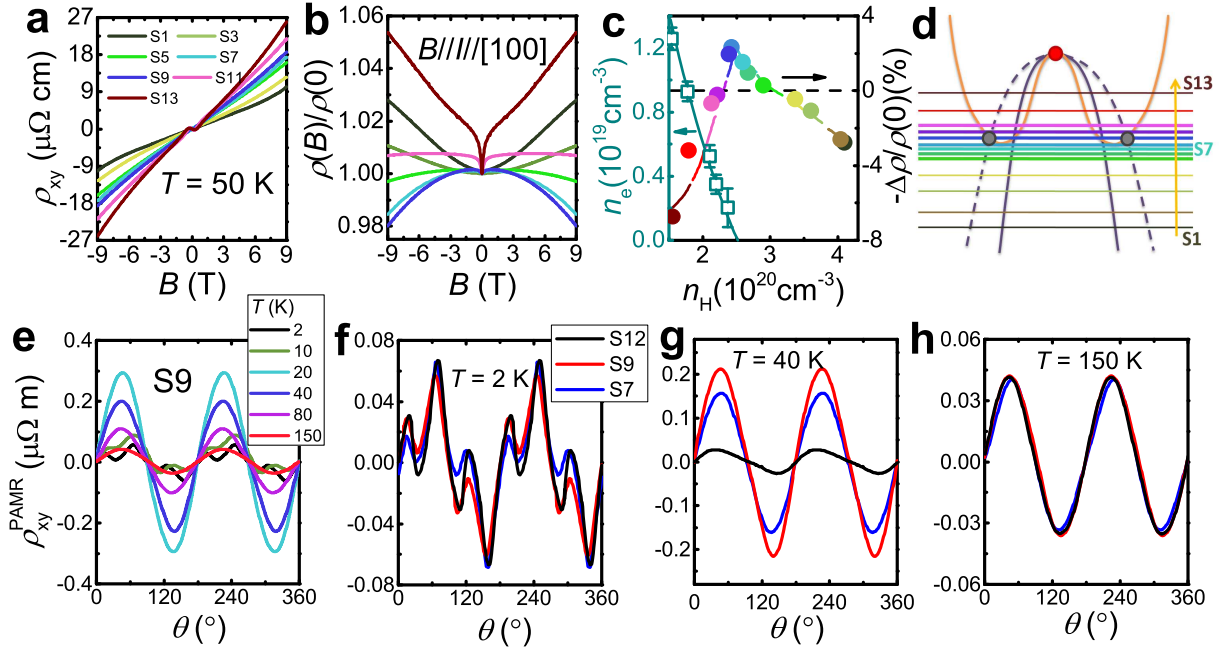}
\end{center}
	\caption{\textbf{Tuning the carrier concentration and transverse resistivity.} \textbf{a}, Hall resistivity, and  \textbf{b}, longitudinal magnetoresistance ($\mathbf{B}\parallel\mathbf{I}$) for different samples, where S11-S13 were Au doped. \textbf{c}, The strength of the negative longitudinal magnetoresistance as $-\Delta\rho/\rho(0)$ [$\Delta\rho=\rho(9{\rm T})-\rho(0)]$, as a function of the hole density $n_{\rm H}$ from fitting the Hall resistivity using the least squares method, where the error bars correspond to the standard errors (Supplementary Fig.~7 and Supplementary Note~3). For samples with two-band behavior, the electron density $n_{\rm e}$ is also shown. \textbf{d}, Illustration of $E_{\rm F}$  for different samples where upon electron doping, $E_{\rm F}$  eventually intersects the electron bands, very close to the triply-degenerate fermion points. Samples S5-S11 (thick lines) show evidence for the chiral anomaly. The transverse resistivity  with a field of 9~T rotated in the plane containing the voltage drop and current, where at $\theta=0$  the field and voltage are parallel, is displayed for \textbf{e}, one sample at various temperatures, and for three different samples at \textbf{f}, 2~K, \textbf{g}, 40~K, and \textbf{h}, 150~K. }
   \label{Fig3}
\end{figure}

\clearpage
\begin{figure}[h]
\begin{center}
 \includegraphics[width=0.8\columnwidth]{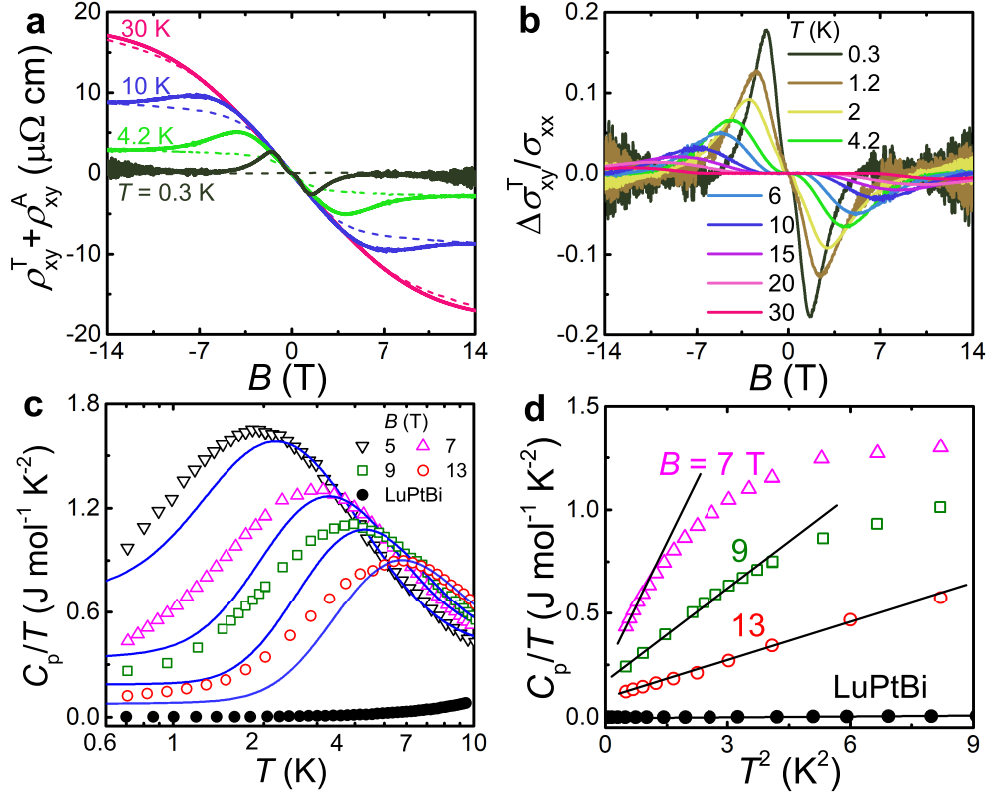}
\end{center}
	\caption{\textbf{Evidence for Weyl fermions in the heavy fermion state of YbPtBi from the anomalous Hall effect and specific heat.} \textbf{a}, Anomalous contribution to the Hall effect of sample S6 obtained from subtracting the ordinary one band Hall resistivity. The Anomalous Hall effect contains two terms, $\rho_{\rm xy}^{\rm A}$ which is proportional to the magnetization, and the topological term $\rho_{\rm xy}^{\rm T}$. The dashed lines show just $\rho_{\rm xy}^{\rm A}$ obtained from analyzing the data together with the measured magnetization (Supplementary Fig.~8 and Supplementary Note~4). \textbf{b} Topological Hall angle $\Delta\sigma_{\rm xy}^{\rm T}/\sigma_{\rm xx}$ as a function of field, after subtracting $\rho_{\rm xy}^{\rm A}$. A clear peak is observed at temperatures up to at least 30~K, giving evidence for the Berry curvature induced by the Weyl points. \textbf{c}, Specific heat  as $C_{\rm p}/T$ at 5~T, 7~T, 9~T, and 13~T where the solid  lines show the results of fitting a  Kondo resonance model  \cite{SCHOTTE197538}. The deviation from the model at low temperatures shows clear evidence for an additional low energy contribution. For comparison the zero-field specific heat of non-magnetic LuPtBi from Ref.~\cite{Mun2013YbPtBi} is also displayed. \textbf{d}, Specific heat in-field, as $C/T$ vs $T^2$, showing that the  low temperature behavior is well accounted for by a $T^3$ dependence (solid lines) which is the expected behavior for Weyl  heavy fermion semimetals  \cite{SiHFEeyl}. }
   \label{Fig4}
\end{figure}

\clearpage

\begin{figure}[h]
\begin{center}
 \includegraphics[width=0.7\columnwidth]{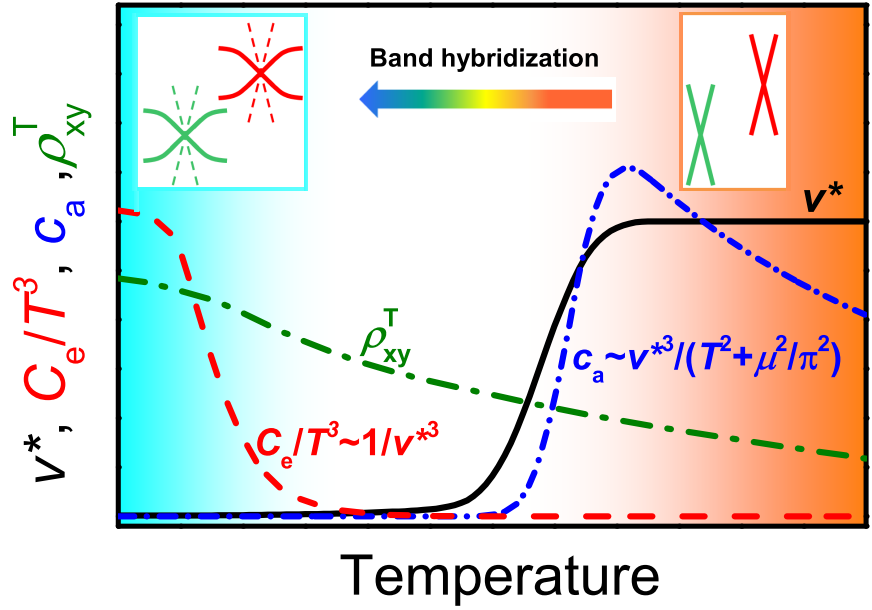}
\end{center}
	\caption{ \textbf{Schematic phase diagram for Weyl fermions in heavy fermion systems.} Illustration of the evolution of the contributions from different experimental probes of Weyl fermions in heavy fermion systems; the chiral anomaly $c_{\rm a}$, electronic specific heat $C_{\rm e}$, and topological Hall effect $\rho_{\rm xy}^{\rm T}$. When the electronic bands in the vicinity of the Weyl points become heavy at low temperatures,  the massive reduction of the effective Fermi velocity $v^*$  leads to the chiral anomaly contribution $c_{\rm a}$ becoming undetectable, yet gives a significant contribution from Weyl nodes to the electronic specific heat $C_{\rm e}$, which is otherwise not detectable in weakly correlated materials. Meanwhile the topological Hall effect $\rho_{\rm xy}^{\rm T}$ which arises from the Berry curvature can be detected both when the $f$-electrons are well localized, as well as deep inside the heavy fermion state. }
   \label{Fig5}
\end{figure}

\end{document}